\documentclass[prd,twocolumn,floats,floatfix,nofootinbib]{revtex4}
\usepackage[dvips]{graphicx}
\usepackage{graphics}
\usepackage{dcolumn}
\usepackage{amssymb}
\usepackage{bm}
\usepackage{amsmath}

\def\spose#1{\hbox to 0pt{#1\hss}}

\def\lta{\mathrel{\spose{\lower 3pt\hbox{$\mathchar"218$}}
     \raise 2.0pt\hbox{$\mathchar"13C$}}}
\def\gta{\mathrel{\spose{\lower 3pt\hbox{$\mathchar"218$}}
     \raise 2.0pt\hbox{$\mathchar"13E$}}}
\newcommand{\be}{\begin{equation}}
\newcommand{\en}{\end{equation}}
\newcommand{\bea}{\begin{eqnarray}}
\newcommand{\ena}{\end{eqnarray}}

\newcommand{\dd}{\mbox{d}}

\newcommand{\ie}{\textsl{i.e.~}}

\newcommand{\eg}{\textsl{e.g.~}}

\newcommand{\GN}{G_{_\mathrm{N}}}

\newcommand{\lP}{\ell_{_\mathrm{Pl}}}
\begin{document}
\title{Bouncing models with a cosmological constant}

\author{Rodrigo Maier,  Stella Pereira, Nelson Pinto-Neto and Beatriz B. Siffert}
\affiliation{ICRA - Centro Brasileiro de
Pesquisas F\'{\i}sicas -- CBPF, \\ Rua Dr. Xavier Sigaud, 150, Urca,
CEP22290-180, Rio de Janeiro, Brazil}

\date{\today}

\begin{abstract}

Bouncing models have been proposed by many authors as a completion, or even as an alternative
to inflation for the description of the very early and dense Universe. However, most bouncing models
contain a contracting phase from a very large and rarefied state, where dark energy
might have had an important role as it has today in accelerating our large Universe. In that
case, its presence can modify the initial conditions and evolution of cosmological
perturbations, changing the known results already obtained in the literature concerning their amplitude
and spectrum. In this paper, we assume the simplest and most appealing candidate for dark energy,
the cosmological constant, and evaluate its influence on the evolution of cosmological perturbations
during the contracting phase of a bouncing model, which also contains a scalar field with a potential
allowing background solutions with pressure and energy density satisfying $p=w\epsilon$, $w$ being a
constant. An initial adiabatic vacuum state can be set at the end of domination by the cosmological constant,
and an almost scale invariant spectrum of perturbations is obtained for $w\approx 0$, which is the usual
result for bouncing models. However, the presence of the cosmological constant
induces oscillations and a running towards a tiny red-tilted spectrum for long wavelength perturbations.

PACS numbers: 98.80.Cq, 04.60.Ds

\end{abstract}

\maketitle

\section{Introduction}

Bouncing models \cite{bergli,bounce} have been widely investigated as a solution of the
singularity problem, and possibly as an alternative to inflation as long as
it can also solve, by its own way, the horizon and flatness problems, and justify
the power spectrum of primordial cosmological perturbations inferred by observations.

In the case where the contracting phase of a regular bouncing model is
dominated by some matter content with a constant ratio between pressure
and energy density, $p/\epsilon = w =$ const., it was shown by many authors, in different
frameworks \cite{fabio1,wands,bozza,fabio2,ppn4,brand}, that this matter content must be dust-like,
perhaps connected to cold dark matter, in order to obtain
a scale invariant spectrum of scalar and tensor cosmological perturbations.

On the other hand, since 1998 \cite{de}, cosmologists were confronted with a
highly unexpected observation: the Universe is presently in a state of
accelerated expansion. This may be caused by the existence of some field
violating the strong energy condition, called dark energy, by a modification
of general relativity at large scales, by the influence of some large scale
inhomogeneities, or simply by a well suited cosmological constant. This
last option is by far the simplest explanation to the present acceleration
of the Universe, although it poses a problem
to quantum field theory on how to accommodate its observed value with vacuum
energy calculations. Anyway, the so called $\Lambda$CDM standard model assumes
that there exists a cosmological constant term in Einstein's equations, which
becomes dynamically important when the typical scale of the Universe has the
size of the present Hubble radius.

In bouncing models without a cosmological constant,
vacuum initial conditions for quantum cosmological
perturbations are set in the far past of the contracting phase, when
the Universe was very big and almost flat, justifying the choice of an
adiabatic Minkowski vacuum in that phase. However, if a cosmological
constant is present, the asymptotic past of bouncing models will
approach de Sitter rather than Minkowski spacetime. Furthermore, the
large wavelengths today become comparable with the Hubble radius in
the contracting phase when the Universe was still slightly influenced by the
cosmological constant.
Hence, the existence of a cosmological constant
can modify the spectrum and amplitude of cosmological perturbations.
Note that this is not a question for inflation because initial conditions
for quantum perturbations and the moment of Hubble radius
crossing in such models take place when the cosmological constant is completely
irrelevant: the Universe is fully dominated by the inflaton field.

The aim of this paper is to investigate this issue in detail in the context
of a Friedmann-Lema\^itre-Robertson-Walker geometry with a cosmological constant,
and a scalar field with potential allowing a constant equation of state $p=w\epsilon$ for the background field,
like the exponential potential in the scenario without cosmological constant. Hence, this paper can
be considered as an extension of
Ref.~\cite{wands} through the introduction of a cosmological constant in the model. Here, as in
Ref.~\cite{wands}, our background scenario is not intended to be a fully realistic description
of the contracting phase of a bouncing model, but to yield a suitable framework to calculate the spectrum 
of linear cosmological perturbations in bouncing models, and to study how it depends on the presence of a
cosmological constant and on the equation of state of the matter content.
In our model, the bounce itself takes place at very short length scales, where the cosmological constant has no role.
Hence, its presence does not modify the evolution
of the background and perturbations in that period, and the descriptions provided in
Refs.~\cite{fabio1,wands,bozza,fabio2,ppn1,ppn2,ppn3,ppn4} can still be considered to be
valid at the bounce.
The main difference is originated from processes much before the bounce, when the initial conditions are
set and the cosmological constant is not irrelevant. In that case, a Minkowski adiabatic vacuum can
only be defined in a precise time domain, i.e., at the end of cosmological constant domination,
but when the Universe was still very big and rarefied. However, even in this time domain, as the length scale
associated with the cosmological constant, given by the present acceleration of the Universe,
is not much bigger than the long wavelengths of physical interest today,
the spectrum of these scales can still be slightly affected by the cosmological constant.
And indeed we will show, analytically and numerically, that the usual result for
bouncing models, namely, that the fluid should satisfy $w\approx 0$ in order to have an
almost scale invariant spectrum of long wavelength perturbations, still holds,
but the presence of the cosmological constant
induces small oscillations and a small running towards a red-tilted spectrum for these scales.

In the next section, we will present the background model and obtain the equations
for the evolution of cosmological perturbations on this background. In section III, we
will discuss the choice of the initial state of the cosmological perturbations
on this background.
In section IV, we will obtain analytically and numerically the power
spectrum of perturbations for the model presented in section III,
and discuss its physical consequences.
We end up with the conclusions.

\section{The background model and the equations for scalar perturbations}

The gravitational action we shall begin with is that of General Relativity with
a cosmological constant, \ie
\begin{equation}
\mathcal{S}_{_\mathrm{GR}}
= -\frac{1}{6\lP^2} \int \sqrt{-g} (R + 2\Lambda) \dd^4 x, \label{action}
\end{equation}
where $\lP=(8\pi\GN/3)^{1/2}$ is the Planck length in natural units
($\hbar=c=1$), and $\Lambda$ is the cosmological constant.

The geometry of the background is given by
the spatially flat homogeneous and isotropic
line element in conformal time:
\begin{equation}
\label{linha-fried}
\dd s^{2}=a^{2}(\eta)(\dd \eta^2 - \delta_{ij}\dd x^{i}\dd x^{j}).
\end{equation}

The matter content of the model is described by a
canonical miminimally coupled scalar field $\varphi$ with Lagrangian

\begin{equation}
\label{R1}
{\cal L}=\frac{1}{2}\varphi_{,\alpha}\varphi^{,\alpha}-U(\varphi),
\end{equation}
where the potential energy density of the scalar field is given by
\begin{equation}
\label{R1p}
U(\varphi)=U_{0}\sinh^2{\Big(\frac{\varphi}{F}\Big)},
\end{equation}

\begin{equation}
\label{R8}
U_0=\frac{3(1-w) H^2_0 \Omega_\Lambda}{16 \pi G}~,~F=\sqrt{\frac{1}{6\pi G (1+w)}},
\end{equation}
and $w$ is a constant. This potential was already studied in Refs.~\cite{ellis1,ellis2} for the
case $w=-1/3$.

In the case of a homogeneous and isotropic background, one can find the scalar field solution

\begin{equation}
\label{R7}
\varphi(t)=\pm \sqrt{\frac{1}{6\pi G (1+w)}}\ln \Big|\tanh\Big[\frac{3(1+w)}{4}\sqrt{\Omega_\Lambda}H_0 t\Big]\Big|,
\end{equation}
and under these conditions the energy density and pressure of the scalar field

\begin{equation}
\label{R2}
\epsilon =\frac{1}{2a^2}{\varphi^{\prime}}^2+U(\varphi)~,~p=\frac{1}{2a^2}{\varphi^{\prime}}^2-U(\varphi),
\end{equation}
satisfy $p=w\epsilon$ (a prime denotes derivative with respect to conformal time).

The Friedmann equations in conformal time read
\begin{equation}
\mathcal{H}^2=\frac{8\pi G}{3}\,a^2\epsilon + a^2\Lambda,
\label{eq_1}
\end{equation}
\begin{equation}
\mathcal{H}^\prime-\mathcal{H}^2=-4\pi Ga^2(\epsilon+p),
\label{eq_3}
\end{equation}
and $\epsilon$ satisfies the conservation equation
\begin{equation}
\label{eq_2}
\epsilon^\prime=-3\mathcal{H}(\epsilon + p),
\end{equation}
where $\mathcal{H}\equiv a'/a$.

In the present situation, where the pressure and energy density of the matter content
satisfy $p=w\epsilon$, with $w$ constant,
the solution for the scale factor in terms of cosmic time $dt=a\ d\eta$ reads,

\begin{widetext}
\begin{equation}
\label{eq_6}
a(t)=a_0\left( \frac{\Omega_{0\omega}}{\Omega_{\Lambda}} \right)^{1/3(1+\omega)} \left[ \sinh \left(- \frac{3\sqrt{\Omega_{\Lambda}} (1+\omega)H_0}{2} t \right)\right]^{2/3(1+\omega)},
\end{equation}
\end{widetext}
where $H_0 = 72$ km s$^{-1}$ Mpc$^{-1}$ is the present Hubble parameter, $\Omega_{0\omega}\equiv \epsilon_0/\epsilon_{\rm crit}$
with $\epsilon_{\rm crit}\equiv 3H_0^2/(8\pi G)$, and $\Omega_{\Lambda}\equiv\Lambda/H_0^2$.
The subscript $0$ indicates the present values of the respective quantities.

Note from Eqs.~(\ref{R1p},\ref{R2}) and $p=w\epsilon$ that $\dot{\varphi}=\sqrt{2(1+w)U_0/(1-w)}\sinh{\varphi/F}$, and
the kinetic energy of the scalar field grows exponentially with $\varphi$, as usually expected for a scalar field in a contracting
phase of a Friedmann model. However, the potential increases in the same way, and that is why  $p = w \epsilon$ is maintained. 
Nevertheless, there is the question about the instability against initial conditions of this tracking between potential 
and kinetic energies of the scalar field 
in order to keep $p = w \epsilon$ in the contracting phase, which is also an issue for Ref.~\cite{wands},
but note that it is not necessary that this tracking must be valid 
all the way to the bounce: here, as in Ref.~\cite{wands}, we are interested in the spectrum of long wavelength perturbations, 
which become comparable
to the Hubble radius in the contracting phase at scales not much smaller than the Hubble radius today. Hence, our calculations 
for the spectrum of long wavelength perturbations should hold even if the tracking $p = w \epsilon$ ceases to be valid at smaller scales. 
This may affect the spectrum of small wavelength perturbations, but it will not affect our forthcoming calculations 
for long wavelength perturbations if the
tracking $p = w \epsilon$ holds at scales of the order of the Hubble radius today, when the contracting Universe is still
very big. Hence, in the same way that the conclusions of 
Ref.~\cite{wands} proved to be valid for more elaborate bouncing models, we expect that the results we will present
later on will also be valid in more elaborate models containing a cosmological constant.

Note also that as $1/\sqrt{\Lambda}$ is of the order of the Hubble radius today, 
then all scales of physical interest had
causal contact in the contracting phase of the model because the particle horizon in that phase
is of the order of $1/\sqrt{\Lambda}$.

The evolution of linear scalar perturbations are described by
the metric

\begin{equation}
\label{perturb}
g_{\mu\nu}=g^{\rm back}_{\mu\nu}+h_{\mu\nu},
\end{equation}
where $g^{\rm back}_{\mu\nu}$ represents the homogeneous and isotropic
cosmological background given in Eq.~(\ref{linha-fried}), and the perturbations
$h_{\mu\nu}$ can be decomposed into
\begin{eqnarray}
\label{perturb-componentes}
h_{00}&=&2a^{2}\phi \nonumber \\
h_{0i}&=&-a^2B_{,i} \\
h_{ij}&=&2a^{2}(\psi\gamma_{ij}-E_{,ij}). \nonumber
\end{eqnarray}

Due to the constraint equations present in the Einstein's equations, the evolution of quantum
perturbations in a classical background is
described by a single quantum field, the gauge invariant Mukhanov-Sasaki variable
defined by (see Ref.~\cite{bookM} for details)

\begin{equation}
\label{MS}
v\equiv a\,\biggl(\delta\varphi+\frac{{\varphi'_{\rm back}}}{\mathcal{H}}\psi\biggr),
\end{equation}
where $\delta\varphi$ is the perturbed scalar field, and $\varphi_{\rm back}$ is its
background solution.

The Mukhanov-Sasaki variable satisfies the equation,

\begin{equation}
\label{eqMS}
v''- \nabla^2 v - \frac{z''}{z}v = 0,
\end{equation}
where

\begin{equation}
\label{z}
z\equiv \frac{a^2\sqrt{4\pi G_N(\epsilon + p)}}{\mathcal{H}}.
\end{equation}

The equations above are not altered by the presence of a cosmological constant,
essentially because Eqs.~(\ref{eq_3},\ref{eq_2}) are not modified by its presence
and because, of course, $\delta\Lambda =0$.

In our choice of units $a$ is dimensionless, hence we will define the dimensionless
conformal time ${\tilde{\eta}}\equiv \eta /R_H$, where $R_H \equiv 1/(a_0 H_0)$ is the co-moving Hubble radius.
From now on we will omit the tilda over $\eta$.
We will also work with the dimensionless comoving wavenumber $k\equiv R_H/\lambda$, where $\lambda$ is the comoving
wavelength of the perturbation modes. The region corresponding to long wavelengths today is
the interval $1<k<10^3$.

Taking the model with scalar field and scale factor
given by Eqs.~(\ref{R7}) and (\ref{eq_6}), respectively, one obtains that

\begin{equation}
\label{z1}
z(t) = -\sqrt{\frac{3(1+w)}{2}}\frac{a(t)}{\cosh(\gamma t)},
\end{equation}
and
\begin{widetext}

\begin{equation}
\label{V}
V(t)\equiv\frac{z''}{z}=\frac{\Omega_{\Lambda}a^2}{a_0^2}\left\{\frac{(1-3w)}{2}\left[\frac{1}{\sinh ^2 (\gamma t)}
-\frac{(1+3\omega)}{2} \right] - \frac{9(1+w)^2}{2\cosh^2(\gamma t)}\right\},
\end{equation}
\end{widetext}
where

\begin{equation}
\label{gamma}
\gamma\equiv\frac{3\sqrt{\Omega_{\Lambda}} (1+\omega)H_0}{2},
\end{equation}
and $a$ is given by Eq.~(\ref{eq_6}).

Solution (\ref{eq_6}) is defined in two domains: $-\infty < t <0$ and $0<t<\infty$. The first one describes a universe contracting
from an asymptotic de Sitter spacetime in the far past to a singularity at $t=0$. The second one describes a universe expanding
from a singularity at $t=0$ to an asymptotic de Sitter expansion in the far future. Around $t=0$, the field dominates the
dynamics, and the cosmological constant is unimportant.
These behaviours can be viewed by taking the limits \eg, in the contracting solution,
$t\rightarrow -\infty $ and $t\rightarrow 0^- $ in Eq.~(\ref{eq_6}).

For $t\rightarrow -\infty $, Eq.~(\ref{eq_6}) yields $a(t)\approx \exp (-\sqrt{\Lambda}t)$. In conformal time
\begin{equation}
\label{conformal}
\eta + \eta_{\infty}=\left(\frac{4}{\Omega_{0\omega}}\right)^{1/3(1+\omega)}
\frac{\exp (\sqrt{\Lambda}t)}{{\Omega_{\Lambda}}^{(1+3\omega)/3(1+\omega)}},
\end{equation}
where $-\eta_{\infty}<\eta \ll0$, and $\eta_{\infty}$ is a positive constant, the scale factor behaves as

\begin{equation}
a(\eta) = \frac{a_0}{\sqrt{\Omega_{\Lambda}}(\eta + \eta_{\infty})}. 
\end{equation}
This is the usual de Sitter behaviour. In this case, the potential
(\ref{V}) reads

\begin{equation}
\label{V1}
V(\eta)\equiv\frac{z''}{z}\approx \frac{9w^2-1}{4(\eta + \eta_{\infty})^2},
\end{equation}
yielding the equation
\begin{equation}
\label{vads}
v''_k + \left[k^2 - \frac{(9w^2-1)}{4(\eta + \eta_{\infty})^2}\right]v_k=0..
\end{equation}

This equation is completely equivalent to an equation for a massive scalar
field in a de Sitter spacetime, with mass given by

\begin{equation}
\label{mass}
m = \frac{3\sqrt{\Lambda}}{2}\sqrt{1-w^2}.
\end{equation}
Its general solution reads

\begin{equation}
\label{solg}
v_k = \sqrt{\eta}\left[b_1(k) H_{\nu}^{(1)}(k(\eta + \eta_{\infty})) + b_2(k) H_{\nu}^{(2)}(k(\eta + \eta_{\infty}))\right],
\end{equation}
where the $H_{\nu}^{(1,2)}$ are Hankel functions of first and second kind, $\nu =3w/2$.
As $k(\eta + \eta_{\infty})\ll 1$, we can write this solution as

\begin{equation}
\label{bds}
v_k \approx c_1(k) (\eta + \eta_{\infty})^{(1+3w)/2} + c_2(k) (\eta + \eta_{\infty})^{(1-3w)/2}.
\end{equation}

For $t\rightarrow 0^- $, or $\eta \rightarrow 0^- $, one obtains from Eq.~(\ref{eq_6}) that
$a(t)\propto t^{2/[3(1+w)]}$ or, in conformal time, $a(\eta)\propto \eta^{2/(1+3w)}$. This is the
usual Friedmann evolution for $p=w\epsilon$ without a cosmological constant. In this regime, we obtain

\begin{equation}
\label{V2}
V(\eta)\equiv\frac{z''}{z}\approx \frac{2(1-3w)}{(1+3w)^2\eta^2}.
\end{equation}
In this situation, $z\propto a$ (see Eq.~(\ref{z1})) and $z''/z = a''/a$.

\begin{figure}
\begin{center}
\includegraphics[scale=0.75]{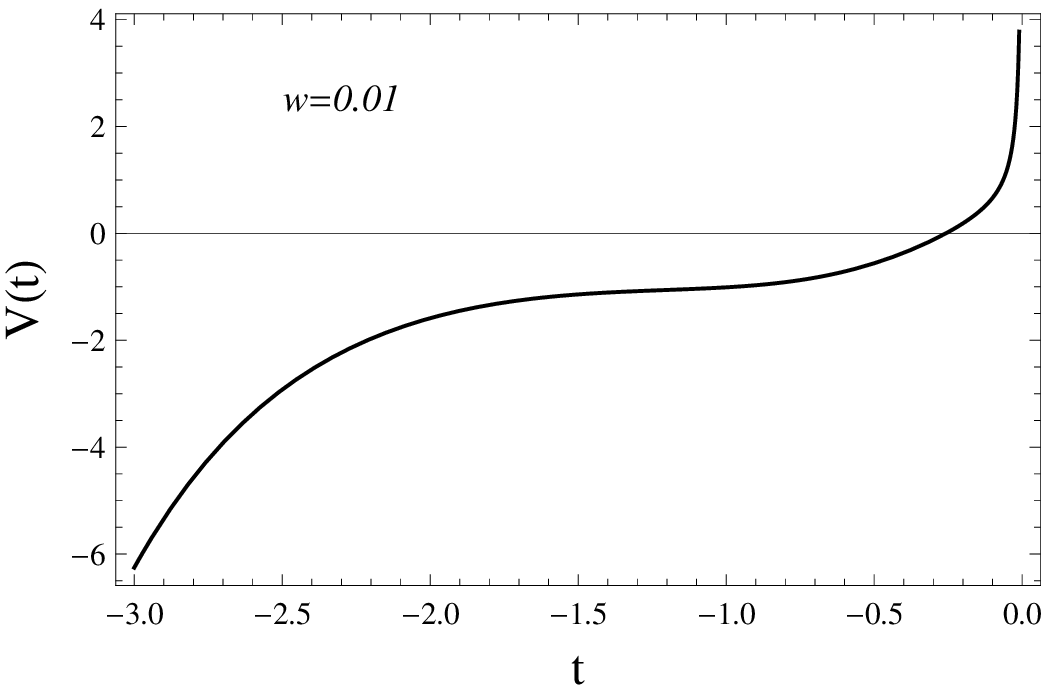}\label{Fig1a}
\includegraphics[scale=0.75]{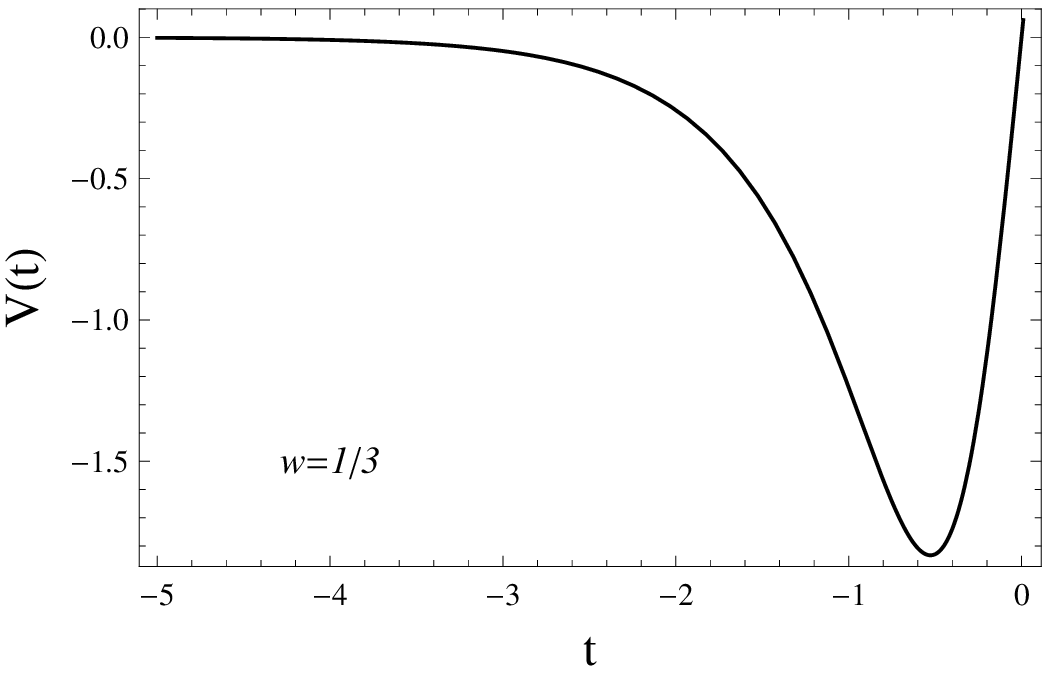}\label{Fig1b}
\includegraphics[scale=0.75]{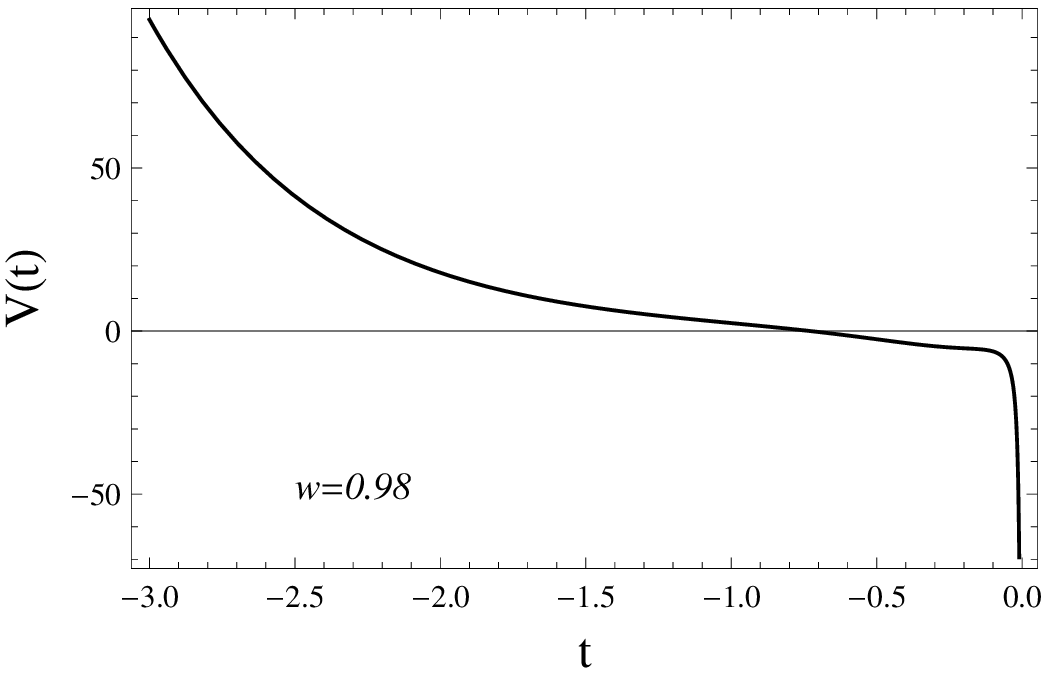}\label{Fig1c}
\caption{Behaviour of the potential $V(t)$ given by Eq.~(\ref{V}) for three different values of $w$.} \label{Fig1}
\end{center}
\end{figure}

Note that the potential $V(t)$ diverges to $\pm \infty$ at the infinity past for $w>1/3$ and $w<1/3$, respectively,
and diverges to $\pm \infty$ near the singularity at $t\propto\eta\approx 0$ for $w<1/3$ and $w>1/3$, respectively.
Hence, it must cross zero in the middle of the line $-\infty <t<0 $. In Fig.~\ref{Fig1} we present the behaviours of the
potential $V(t)$ for $t<0$ in the cases $w<1/3$, $w=1/3$ and $w>1/3$.

Our idea is that the singularity at $t=0$ separating the contracting and expanding solutions can be
eliminated through some new physics which produces a regular bounce connecting these two phases.
As in the region around $t=0$ the cosmological constant is unimportant, one can evoke
the bounce descriptions provided \eg, in Refs.~\cite{fabio1,wands,bozza,fabio2,ppn1,ppn2,ppn3,ppn4}.
For instance,
the quantum cosmological bounces with a perfect fluid studied in Refs.~\cite{ppn1,ppn2,ppn3,ppn4}
present a regular scale factor given by

\begin{equation}
\label{at} a(T) = a_b
\left[1+\left(\frac{T}{T_b}\right)^2\right]^\frac{1}{3(1-\omega)} ,
\end{equation}
where $dT = a^{1-3w} d\eta$, and $a_b$ and $T_b$ are positive constants. Note that for $|T| \gg T_b$, this solution approaches
the classical Friedmann solution for a perfect fluid given by the limit
$t\rightarrow 0^- $ of Eq.~(\ref{eq_6}): $a(\eta)\propto \eta^{2/(1+3w)}$.
Hence, the scale factor (\ref{eq_6}) can be smoothly connected to the scale factor (\ref{at}).

\begin{figure}
\includegraphics[scale=0.75]{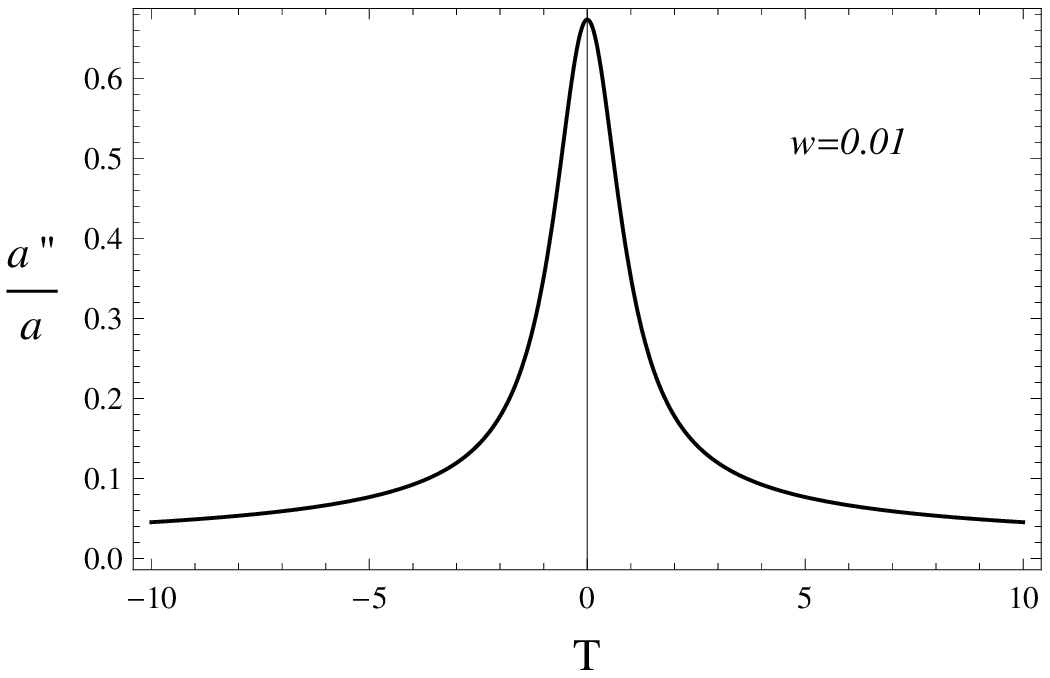}\label{Fig2a}
\includegraphics[scale=0.75]{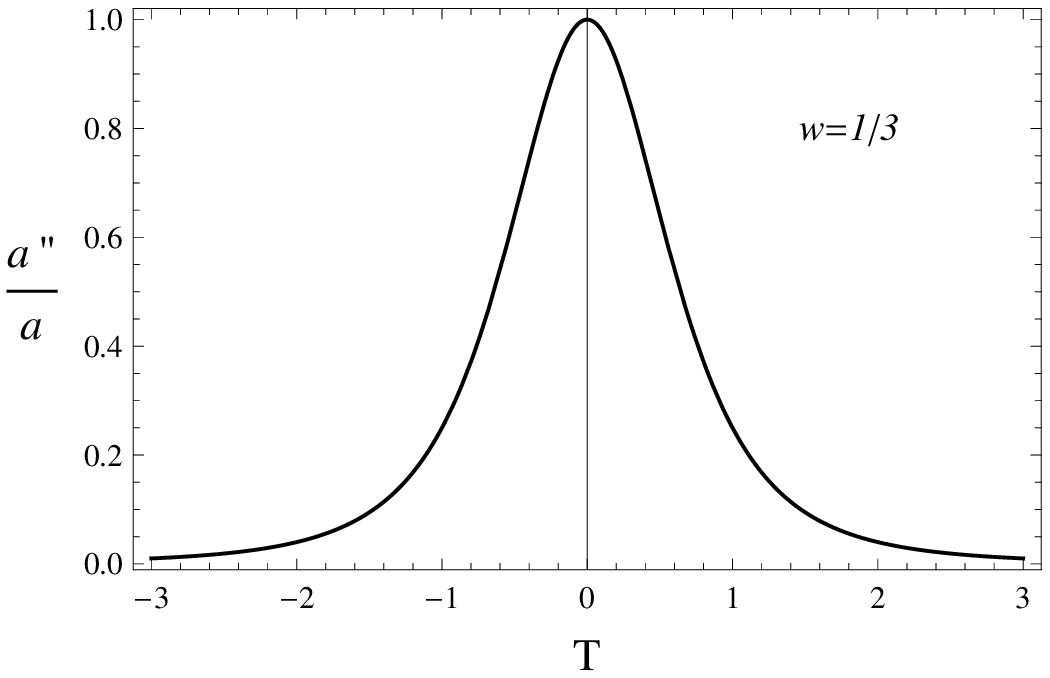}\label{Fig2b}
\includegraphics[scale=0.75]{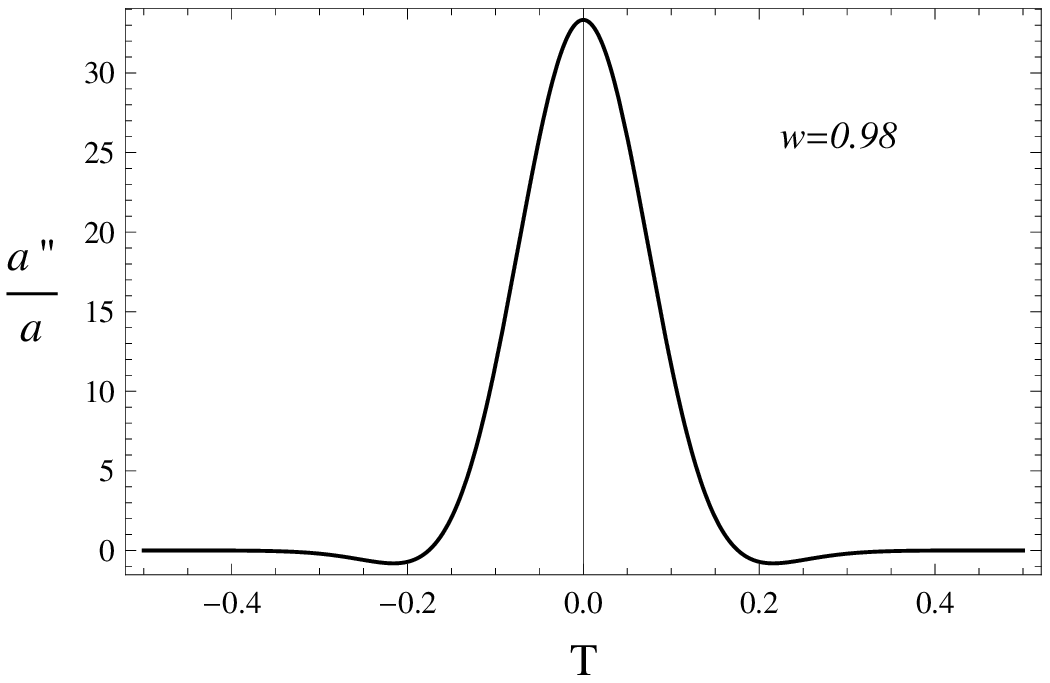}\label{Fig2c}
\caption{Behaviour of the quantum bouncing potential given by Eq.~(\ref{at}) for three different values of $w$.} \label{Fig2}
\end{figure}

It was shown in Ref.~\cite{ppn3} that the potential present
in the equations for the perturbations around
these quantum bounces reads

\begin{equation}
\label{eqQBV}
V(T)\equiv \frac{a''}{a}=
\frac{2\,\,a_b^{3(1-w)}}{3(1-w)T_b^2}\left[\frac{1}{a^{1+3w}}-
\frac{2}{3}\left(\frac{T}{T_b}\right)^2\frac{a_b^{3(1-w)}}{1-w}a^{-4}\right],
\end{equation}
where Eq.~(\ref{at}) has been used.
Their shapes are presented in Fig.~\ref{Fig2} for the cases $w<1/3$, $w=1/3$ and $w>1/3$,
and they tend to the potential $V(t)$ of Eq.~(\ref{V2}) for $\gamma t << 1$
in the limit $|T| \gg T_b$, but they do not diverge in $\eta=0$.
Hence, when these quantum effects become important near $\eta=0$ inducing the bounce,
the two disjoint parts of the classical potentials presented in Fig.~\ref{Fig1} corresponding to the contracting
and expanding classical universes separated by a singularity can now be softly connected
with the potentials presented in Fig.~\ref{Fig2}. Then one can evolve smoothly the perturbations from the
contracting phase to the expanding phase, and calculate their properties in the present era.
For other regular bouncing models, the situation must be similar, although
somewhat more intrincate in the case there is an extra field which induces the bounce.

In order to accomplish this program, one must set the initial conditions for the perturbations
in the past contracting phase. Without the cosmological constant, the Universe tends,
in the far past, to Minkowski spacetime, where the potentials become null (see Fig.~\ref{Fig2}). Hence, an adiabatic
Minkowski vacuum can be prescribed there. In the presence of the cosmological constant, neither
the Universe tends to Minkowski spacetime
in the far past (in fact, it tends to de Sitter spacetime), nor the potential becomes null there
(see Fig.~\ref{Fig1}, except for $w=1/3$, which is not physically interesting because it does not
yield an almost scale invariant spectrum of perturbations). However, as we have shown
above, the potential crosses zero somewhere in the middle of its evolution (which coincides
with the moment when physically interesting long wavelengths perturbation modes
become smaller than the Hubble radius),
and perhaps one could define an adiabatic Minkowski vacuum there. We will show in the next section
that this is indeed possible for the scalar field model presented above.

\section{The choice of initial state for the quantum perturbations}

In this section we will check whether an adiabatic Minkowski vacuum can indeed be prescribed in the
time interval when the potential (\ref{V}) is negligible and the Universe is starting to be dominated by the field. In this regime the scale factor is approaching the form $a(\eta)\propto \eta^{2/(1+3w)}$. Taking $\Omega_{0\omega}\approx 0.3$ and $\Omega_{\Lambda}\approx 0.7$, the zero of this potential occurs when $y=y_V \equiv\gamma t_V$ (see Eq.~(\ref{gamma}) for the definition of $\gamma$). As $y_V$ depends on
the $\omega$ parameter, we will consider the following values for $\omega<1/3$:
\begin{eqnarray}
\label{eqyvn}
\nonumber
y_V(\omega\approx 0) &\approx& -0.34~,
\nonumber
\\ y_V(\omega\approx 1/8) &\approx& -0.23~,
\nonumber
  \\ y_V(\omega\approx 1/4) &\approx& -0.13.
  \nonumber
\end{eqnarray}
Around these points, one can numerically approximate Eq.~(\ref{eqMS}), now expressed in terms
of the modes $v_k(\eta)$, through
\begin{equation}
\label{eqMSk}
\frac{d^2 v_k}{d x^2} + (k^2 + \beta x) v_k = 0,
\end{equation}
where
\begin{equation}
\label{eqMSkl}
\beta \approx -\frac{3(1+w)\sqrt{\Omega_{\Lambda}}}{2} \Big(\frac{a}{a_0} \frac{dV}{dy} \Big)\Big|_{y_V},
\end{equation}
yielding
\begin{eqnarray}
\label{eqMSkl2}
\nonumber
\beta(\omega\approx 0) &\approx& -1.05~,
\nonumber
\\ \beta(\omega\approx 1/8) &\approx& -1.65~,\nonumber
  \\ \beta(\omega\approx 1/4) &\approx& -2.31.
  \nonumber
\end{eqnarray}
and $x\equiv \eta - \eta_V $, with $\eta_V$ being the dimensionless conformal time corresponding
to $y_V$ defined above.

It is important to remark the dependence of $\beta$ with $\Omega_{\Lambda}$ by looking
at Eqs.~(\ref{eq_6}) and (\ref{V}). Noting that $y_V$ is independent of $\Omega_{\Lambda}$,
and as $\Omega_{0w} + \Omega_{\Lambda}=1$, one gets

\begin{equation}
\label{beta-lambda}
\beta = \frac{\Omega_{\Lambda}^{(1+3w)/[2(1+w)]}\left(1-\Omega_{\Lambda}\right)^{1/(1+w)}}
{0.7^{(1+3w)/[2(1+w)]}0.3^{1/(1+w)}}\beta_{0.7},
\end{equation}
where $\beta_{0.7}$ are the values of $\beta$ for $\Omega_{\Lambda}=0.7$.
One can see that $\beta \rightarrow 0$ as $\Omega_{\Lambda} \rightarrow 0$.

Note also that, although $y_V$ is independent of $\Omega_{\Lambda}$, $\eta_V$ depends on $\Omega_{\Lambda}$ as

\begin{equation}
\label{etaV-lambda}
\eta_V = \frac{0.7^{(1+3w)/[6(1+w)]}0.3^{1/[3(1+w)]}}
{\Omega_{\Lambda}^{(1+3w)/[6(1+w)]}\left(1-\Omega_{\Lambda}\right)^{1/[3(1+w)]}}\eta_{V(0.7)},
\end{equation}
where, again, $\eta_{V(0.7)}$ are the values of $\eta_V$ for $\Omega_{\Lambda}=0.7$.
One can see now that $\eta_V \rightarrow \infty$ as $\Omega_{\Lambda} \rightarrow 0$, as expected.
In this last calculation, we have assumed that the field dominates at $\eta_V$.

The adiabatic vacuum is defined by the solution

\begin{equation}
\label{av}
v_k(x) = \frac{1}{2{[\Omega_k (x)]}^{1/2}}\exp\left[-i \int^x \Omega_k (x')dx'\right],
\end{equation}
where $\Omega_k(x)$ must satisfy the equation

\begin{equation}
\label{omega}
\Omega_k^2 = f_k^2-\frac{1}{2\Omega_k}\frac{d^2 \Omega_k}{d x^2}+\frac{3}{4\Omega_k^2}{\left(\frac{d\Omega_k}{d x}\right)}^2,
\end{equation}
and $f_k^2\equiv k^2 + \beta x$.

Order by order, one has:

\begin{eqnarray}
\label{omegas}
(\Omega_k^{(0)})^2 &=& f_k^2; \;\;\;\; (\Omega_k^{(2)})^2 = f_k^2\left(1+\frac{5}{16}\frac{\beta^2}{f_k^6}\right)\nonumber\\
(\Omega_k^{(4)})^2 &=& f_k^2\left[1+\frac{5}{16}\frac{\beta^2}{f_k^6}-\frac{490}{256}{\left(\frac{\beta^2}{f_k^6}\right)}^2\right],
\end{eqnarray}
where the upper indices $(n)$ denote the order of the approximation.
Hence, an adiabatic Minkowski vacuum can be obtained if the parameter expansion $\beta^2/f_k^6$
satisfies $\beta^2/f_k^6<<1$.
In fact, as $\beta$ is of order unity, $x<<1$,
and the long wavelengths of physical interest satisfy $1<k<10^3$, the condition
$\beta^2/f_k^6 \approx \beta^2/k^6 << 1$ is satisfied.
Note that for the largest scales ($k$ approaching $1$),
deviations from the Minkowski vacuum become more significant, and one should expect modifications
against the standard results.

This problem can be presented under another point of view. A Minkowski vacuum can be defined for quantum perturbations
with wavelengths much smaller than the Hubble radius, defined by ${\cal{R}}_H(t)\equiv 1/H(t)$. From Eq.~(\ref{eq_6}), one obtains that

\begin{equation}
\label{sound-horizon}
{\cal{R}}_H(t) = \sqrt{\frac{1}{\Lambda}}\tanh(-y).
\end{equation}

One has to compare this quantity with the physical wavelength $\lambda_{\rm phys} = a\lambda$ which, from Eq.~(\ref{eq_6}),
reads

\begin{equation}
\label{compare}
\lambda_{\rm phys}=\lambda_{\rm phys}^{\rm now}\left( \frac{\Omega_{0\omega}}{\Omega_{\Lambda}} \right)^{1/3(1+\omega)}
\sinh^{2/3(1+\omega)}(-y).
\end{equation}
The maximum value of ${\cal{R}}_H$, at $t\rightarrow -\infty$, is ${\cal{R}}_H^{\rm max}(t) = \Lambda^{-1/2}$, while
$\lambda_{\rm phys}$ diverges there. Eventually, they can be comparable at some time in the contracting phase.
As in the case this is true one expects to obtain a similar spectrum as the one obtained in bouncing models
without a cosmological constant, we will concentrate on the case $0<w<<1$, which yields an almost scale invariant
spectrum. The quantities defined in Eqs.~(\ref{sound-horizon}) and (\ref{compare}) are comparable when

\begin{equation}
\label{condition}
\Omega_{0\omega}^{1/3}\Omega_{\Lambda}^{1/6}\frac{\lambda_{\rm phys}^{\rm now}}{{\cal{R}}_H^{\rm now}}
\approx\frac{\sinh^{1/3}(-y)}{\cosh(-y)},
\end{equation}
where ${\cal{R}}_H^{\rm now}$ is the Hubble radius today. As $\sinh^{1/3}(-y)/\cosh(-y) < 0.73$,
$\Omega_{0 m}^{1/3}\Omega_{0\Lambda}^{1/6}\approx 0.63$, and
$10^{-3} < \lambda_{\rm phys}^{\rm now}/{\cal{R}}_H^{\rm now} < 1$, this equality can be achieved
for some finite domain of $y$.

Note also from
Eq.~(\ref{condition}) that this domain interval of $y$ could be extended to large values of $|y|$
if $\Omega_{\Lambda}$ were much smaller than our prescribed values.
This can also be seen from the analysis coming from the potential, where a smaller
$\Omega_{\Lambda}$ would result in a smaller $\beta$ in Eq.~(\ref{omegas}), and an adiabatic vacuum could
also be achieved for smaller values of $k$.

\section{Spectrum of quantum cosmological perturbations} \label{sectionIV}

Let us now calculate the spectrum of quantum cosmological perturbations for this scenario.
In Section III, we have shown that an adiabatic Minkowski vacuum, for the case of
a canonical scalar field, could
be prescribed in the time interval where the potential becomes negligible
and the Universe is starting to be dominated by the field. Hence,
substituting the zero order term ($\Omega^{(0)}_k$) given in Eq.~(\ref{omegas})
in the solution (\ref{av}) of Eq.~(\ref{eqMSk}), we obtain
\begin{equation}
\label{eqR12}
v_k(x)\approx \frac{1}{2(k^2+\beta x)^{1/4}} \exp\Big[-\frac{2ik^3}{3\beta}\Big(1+\frac{\beta x}{k^2}\Big)^{3/2}\Big],
\end{equation}
and the initial conditions are given by
\begin{equation}
\label{eqR13}
v_k(0)\approx \frac{1}{2\sqrt{k}} \exp\Big(-\frac{2ik^3}{3\beta}\Big)~,
\end{equation}
\begin{equation}
\label{eqR14}
\frac{dv_k}{dx}\Big|_{x=0} \approx -v(0)\Big(\frac{\beta}{4k^2}+ik\Big).
\end{equation}
\par

We have, therefore, to solve

\begin{equation}
\label{eq69}
v_k'' + \left(k^2 - \frac{z''}{z}\right) v_k = 0,
\end{equation}
with initial conditions given by Eqs.~(\ref{eqR13}) and (\ref{eqR14}).

\begin{figure}
\begin{center}
\includegraphics[scale=0.95]{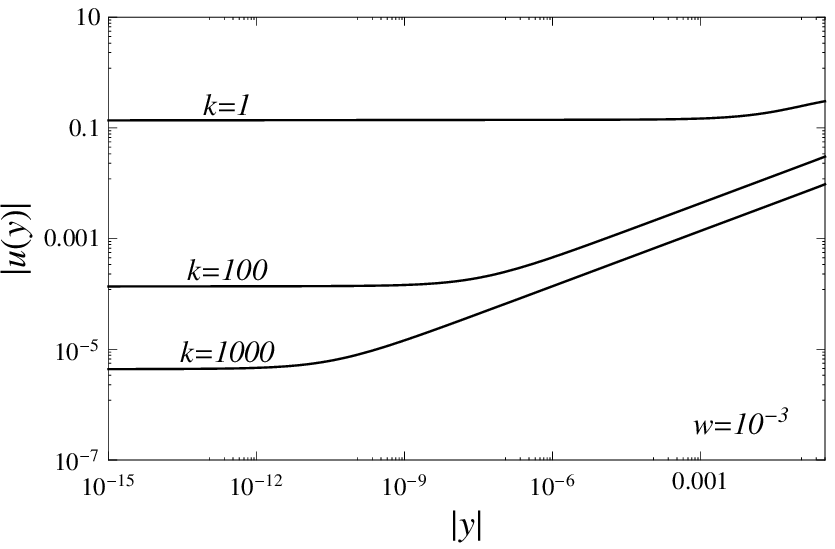}\label{Fig3a}
\includegraphics[scale=0.95]{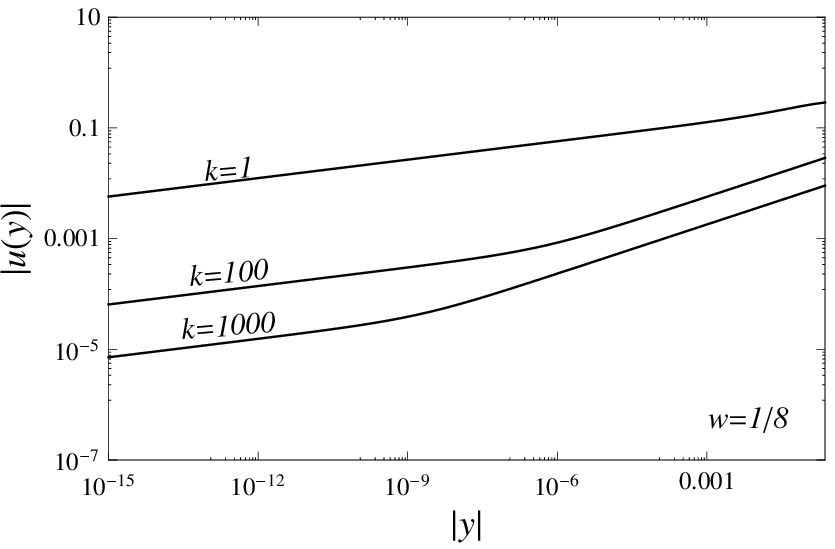}\label{Fig3b}
\includegraphics[scale=0.95]{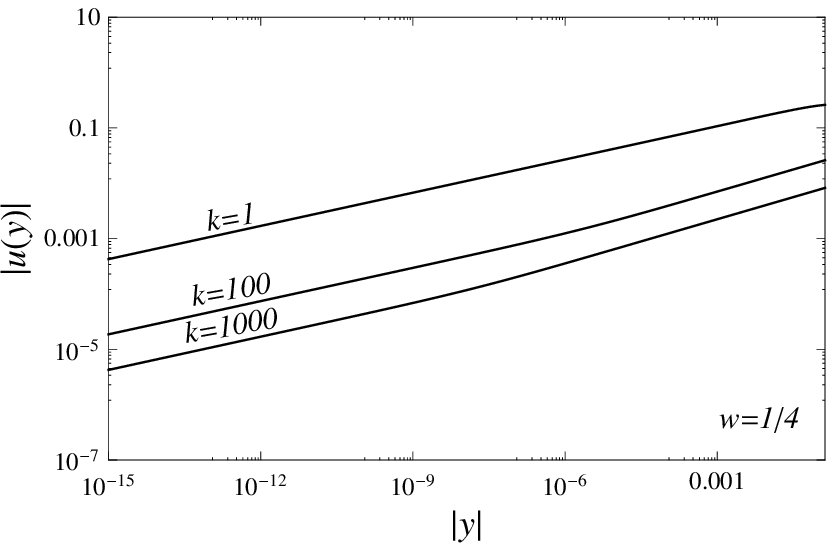}\label{Fig3c}
\caption{Numerical results of Eq. (\ref{numeric}) for $w=10^{-3},1/8$ and $1/4$. Each curve shows $u_k(y)$ as a function of $y$ for different values of $k$, as indicated in the graphs.} \label{Fig3}
\end{center}
\end{figure}

\begin{widetext}
We calculated numerically the solution of Eq.~(\ref{eq69})
by changing the time variable from $\eta$ to $y$, defining a new function $u_k \equiv a^{1/2}v_k$,
and setting the above initial conditions at $y_{\rm ini}=y_V$.
Taking the potential (\ref{V}), the transformed
equation reads

\begin{equation}
\label{numeric}
\frac{d^2 u_k}{dy^2}+\left\{\frac{4k^2}{9(1+w)^2[\Omega_{\Lambda}^{(1+3w)/2}\Omega_{0 w}]^{2/[3(1+w)]}\sinh^{4/[3(1+w)]}(-y)} -    \frac{w^2}{(1+w)^2}+\frac{w}{(1+w)^2\sinh^2(y)} + \frac{2}{\cosh^2(y)}\right\}u_k=0.
\end{equation}
The results are shown in Fig. \ref{Fig3}.

The solutions of equation (\ref{eq69})
can be expanded in powers of $k^2$ according to
the formal solution~\cite{MFB}
\begin{equation}
\frac{v}{z}  \simeq A_1(k)\biggl[1 - k^2 \int^\eta \frac{\dd\bar
  \eta}{z^2\left(\bar \eta\right)} \int^{\bar{\eta}}
  z^2\left(\bar{\bar{\eta}}\right)\dd\bar{\bar{\eta}}+ ...\biggr] +
  A_2(k) \biggl[\int^\eta\frac{\dd\bar{\eta}}{z^2\left(\bar \eta\right)} - k^2
  \int^\eta \frac{\dd\bar{\eta}}{z^2\left(\bar \eta\right)} \int^{\bar{\eta}} z^2\left(\bar{\bar \eta}\right)
  \dd\bar{\bar{\eta}} \int^{\bar{\bar{\eta}}}
  \frac{\dd\bar{\bar{\bar{\eta}}}}{z^2\left(\bar{\bar{\bar \eta}}\right)} + ...\biggr], \label{solform}
\end{equation}
\end{widetext}
In Eq.~(\ref{solform}),
the coefficients $A_1(k)$ and $A_2(k)$ are two constants depending only on
the wavenumber $k$ through the initial conditions.

Since expansion (\ref{solform}) is valid at all times during contraction, the $A_1(k)$ and $A_2(k)$
dependencies coming from the initial conditions
hold when the Universe enters the field dominated phase just before performing
the bounce. From then on, the evolution is guided by the particular physics of the bounce.
For instance, in the quantum bounce with potentials shown in Fig.~\ref{Fig2},
everything goes as described in Ref.~\cite{ppn4}, with power spectrum in
the expanding phase given by

\begin{equation}
\label{PS} \mathcal{P} \propto  k^3 |A_2(k)|^2,
\end{equation}
and spectral index

\begin{equation}
\label{si} n_{_\mathrm{S}} = 1+\frac{d\ln(\mathcal{P})}{d\ln(k)}.
\end{equation}

This is a general feature of bouncing models: the spectrum $A_2(k)$ of the growing mode of the contracting
phase, which will be the decaying mode of the expanding phase after the bounce,
is transferred to the growing mode of the expanding phase and dominates over $A_1(k)$. Hence,
to find the spectrum after the bounce one has to obtain $A_2(k)$, which, in the case without
cosmological constant, reads $A_2(k)\propto k^{\frac{3\left(\omega-1\right)}{2\left(3\omega+1\right)}}$
(see e.g., Ref.~\cite{ppn4}), yielding the spectral index

\begin{equation}
\label{indexS2} n_{_\mathrm{S}} = 1+\frac{12\omega}{1+3\omega},
\end{equation}
with $w\approx 0$ for an almost scale invariant spectrum.

In order to analytically predict the behaviour of the coefficient $A_2(k)$ for the
case with a cosmological constant, we have taken two
approximate solutions of Eq.~(\ref{eq69}) close to the point
where the potential vanishes, where it has the form given in Eq.~(\ref{eqMSk}),
and matched them at a point $x_\ast=(\eta_\ast - \eta_V)$, where $0<x_\ast << 1$.

The first approximate solution comes from Eq.~(\ref{eqR12}), and is given by

\begin{equation}
\label{eqR15}
v_k(x_\ast)\approx \frac{1}{2\sqrt{k}}\Big(1-\frac{\beta x_\ast}{4k^2}\Big) \exp\Big[-i\Big(\frac{2k^3}{3\beta}+k x_{\ast}\Big)\Big],
\end{equation}
while its first derivative reads
\begin{eqnarray}
\label{eqR16}
\nonumber
\frac{dv_k}{dx}\Big|_{x=x_\ast} \approx &-&\frac{1}{2\sqrt{k}} \Big[\frac{\beta}{4k^2}+ik\Big(1+\frac{\beta x_\ast}{4k^2}\Big)\Big] \nonumber
  \\ &\times& \exp\Big[-i\Big(\frac{2k^3}{3\beta}+k x_{\ast}\Big)\Big].
\end{eqnarray}
\par
The second approximate solution comes from the remark that at $\eta_\ast$ and afterwards,
up to the bounce phase, the evolution of the background is dominated by the scalar field,
where the potential approaches the form given in Eq.~(\ref{V2}), yielding the solution

\begin{eqnarray}
\label{eqR16_5}
v_k(\eta)= \sqrt{\eta} \left[C_1(k)H^1_{\nu}(k\eta)+C_2(k)H^2_{\nu}(k\eta)\right],
\end{eqnarray}
where $\nu\equiv 3(1-w)/2(1+3w)$. In the domain $|k\eta_\ast|>>1$ this solution
reads
\begin{equation}
\label{eqR17}
v_k(\eta_\ast)\approx \frac{B_1(k)}{2\sqrt{k}} \Big[1-\frac{\alpha_1}{2ik\eta_\ast}\Big]+\frac{B_2(k)}{2\sqrt{k}} \Big[1+\frac{\alpha_1}{2ik\eta_\ast}\Big],
\end{equation}
and
\begin{equation}
\label{eqR18}
\frac{dv_k}{dx}\Big|_{\eta=\eta_\ast} \approx \frac{1}{2\sqrt{k}}\Big\{ B_1(k) \Big[ik-\frac{\alpha_1}{2\eta_\ast}\Big]
 - B_2(k) \Big[ik+\frac{\alpha_1}{2\eta_\ast}\Big] \Big\},
\end{equation}
where
\begin{eqnarray}
\label{eqR19}
B_1(k)&\equiv& 2 C_1(k) \exp\Big[i\Big(k\eta_\ast-\frac{\pi\nu}{2}-\frac{\pi}{4}\Big)\Big],
  \\ B_2(k)&\equiv& 2 C_2(k) \exp\Big[-i\Big(k\eta_\ast-\frac{\pi\nu}{2}-\frac{\pi}{4}\Big)\Big],
  \\ \alpha_1&\equiv&\frac{\Gamma(\nu+3/2)}{\Gamma(\nu-1/2)}=\frac{2(1-3w)}{(1+3w)^2}.
\end{eqnarray}
Performing the matching between Eqs.~(\ref{eqR15},\ref{eqR16}) and
Eqs.~(\ref{eqR17},\ref{eqR18}) at $\eta_{\ast}$,
one gets for $C_1(k)$ and $C_2(k)$,
\begin{equation}
\label{eqR20}
C_1(k)\approx\left(\frac{i\beta}{16k^3}-\frac{\beta x_{\ast}}{8k^2}\right)\exp\Big[-i\Big(\frac{2k^3}{3\beta}+k x_{\ast}+\alpha_2\Big)\Big]
\end{equation}
and
\begin{eqnarray}
\label{eqR21}
C_2(k)&\approx& \frac{1}{2}\Big(1+\frac{i\alpha_1}{2k\eta_\ast}-\frac{i\beta}{8k^3}\Big)\nonumber
  \\ &\times&\exp\Big[-i\Big(\frac{2k^3}{3\beta}+k x_{\ast}-\alpha_2\Big)\Big]~,~
\end{eqnarray}
where
\begin{equation}
\label{eqR22}
\alpha_2\equiv k\eta_\ast-\frac{\pi\nu}{2}-\frac{\pi}{4} .
\end{equation}

In the limit $k\eta\rightarrow 0^-$, just before the new physics which generates the bounce,
the solution $v_k(\eta)$ approximately reads
\begin{eqnarray}
\label{eqR23}
v_k(\eta) &\equiv & A_1(k) \eta^{1/2+\nu} + A_2(k) \eta^{1/2-\nu} \nonumber \\
&\approx& \sqrt{\eta} \Bigg\{\left(\frac{k\eta}{2}\right)^{\nu}\frac{1}{\Gamma(\nu+1)}\Bigg[C_1(k)+C_2(k)+
\nonumber  \\&+& i \left[C_1(k)-C_2(k) \right]\cot(\nu\pi) \Bigg]+\nonumber
  \\&+&\left(\frac{k\eta}{2}\right)^{-\nu}
\frac{i\left[C_2(k)-C_1(k)\right]}{\Gamma(1-\nu)\sin(\nu\pi)}\Bigg\},
\end{eqnarray}
and the coefficient $A_2(k)$ of the growing mode of the contracting
phase is given by
\begin{eqnarray}
\label{eqR24}
&&|A_2(k)|^2 \approx  \frac{1}{4}\Big(\frac{k}{2}\Big)^{-2\nu}\frac{1}{[\Gamma(1-\nu)\sin(\nu\pi)]^2} \nonumber
  \\ &\times & \Big [1+\frac{\beta x_{\ast}}{2k^2}\cos(2\alpha_2)+\frac{\alpha^2_1}{4k^2\eta^2_{\ast}}-
  \frac{\beta}{4k^3}\sin(2\alpha_2)\Big].
\end{eqnarray}

Caculating the spectral index as defined in Eq.~(\ref{si}), we find

\begin{eqnarray}
\label{sifinal}
&&n_S = 1+\frac{12w}{1+3w} - \frac{\beta x_{\ast}\eta_{\ast}}{k}\sin(2\alpha_2)\nonumber \\ &-&
\frac{2}{k^2}\Big [\frac{\alpha^2_1}{4\eta^2_{\ast}} + \frac{\eta_{\ast}\beta}{4}(1+2x_{\ast})\cos(2\alpha_2)
+\frac{\eta_{\ast}\beta}{4}\cos(2\alpha_2)\Big] \nonumber \\
&+& \frac{3\beta}{4k^3}\sin(2\alpha_2).
\end{eqnarray}

Substituting the parameters  $w\approx 0$, $\beta \approx -1.05$, $|\eta_{\ast}| \approx |\eta_V| \approx 2.19$, $x_{\ast}<<1$, $\alpha_1\approx 2$, $\alpha_2\approx k\eta_\ast-\pi$, and $1<k<10^3$ in Eq.~(\ref{sifinal}), we obtain an almost scale invariant spectrum.
Besides the usual $12w/(1+3w)$ result, there are additional terms in Eq.~(\ref{sifinal})
inducing a running red-tilted spectrum and oscillations, both decreasing with $k$. Note that for
a vanishing cosmological constant we have $\beta\approx 0$ and $|\eta_{\ast}|\rightarrow\infty$.
In this case, the extra terms in Eq.~(\ref{sifinal}) disappear and $n_S \rightarrow 1$ even for small values of $k$.

In order to check numerically this analytic calculation, we took the following steps:
from the numerical solutions $u_k=a^{1/2}v_k$ presented in Fig.~\ref{Fig3}, we obtained $v_k$, evaluated it at very
small $y$ ($y\approx -10^{-15}$), expressed the result in conformal time (whose relation
with $y$ is trivial at field domination), multiplied the result by $\eta^{-\nu-1/2}$ (see Eq.~(\ref{eqR23})),
and differentiated the final result with respect to $\eta$ in order to isolate $A_2(k)$.
The results are shown in Figs.~\ref{Fig4} and~\ref{Fig5}.

\begin{figure}
\begin{center}
\includegraphics[scale=1]{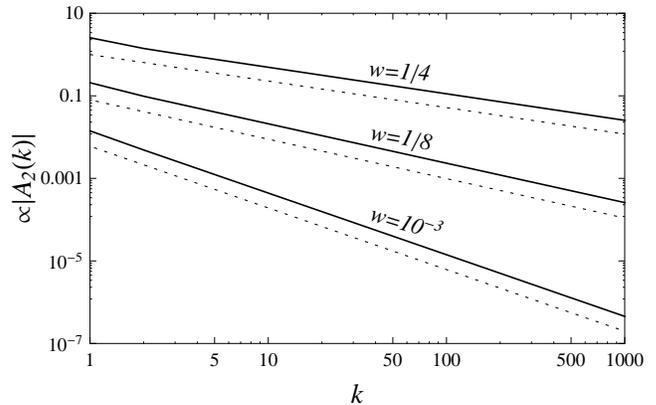}\label{Fig4}
\caption{Numerical results for the behaviour of $|A_2(k)|$ for $w=10^{-3},1/8$ and $1/4$ evaluated at $y=-10^{-15}$. The solid lines show the numerical results and the dotted ones show, for comparison, a curve proportional to $k^{\frac{3\left(\omega-1\right)}{2\left(3\omega+1\right)}}$.} \label{Fig4}
\end{center}
\end{figure}

\begin{figure}
\begin{center}
\includegraphics[scale=0.85]{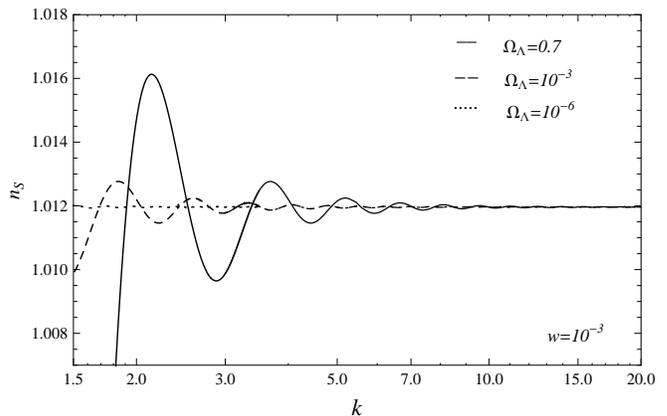}\label{Fig5}
\caption{Numerical results for $n_S(k)$ evaluated at $y=-10^{-15}$, obtained using $w=10^{-3}$. The solid line indicates the result obtained using $\Omega_{\Lambda}=0.7$, the dashed line for $\Omega_{\Lambda}=10^{-3}$ and the dotted line for $\Omega_{\Lambda}=10^{-6}$. 
Note that the oscillations become smaller for
smaller $\Omega_\Lambda$, showing that they are due to the presence of the cosmological constant. This result is in agreement to Eq.~(\ref{sifinal}).} \label{Fig5}
\end{center}
\end{figure}

It can be seen that $A_2(k)$ and $n_S$ follow the predicted behaviour, $A_2(k)\propto k^{3(w-1)/2(1+3w)}$
and $n_S \approx 1+\frac{12w}{1+3w}$,
for $k\gtrsim 1$. We have also verified that small oscillations and a red-tilted running with amplitudes decreasing with $k$
are superimposed to the power law overall behaviour, as predicted in Eqs.~(\ref{eqR24}) and (\ref{sifinal}).
Note from Fig. 5
that the oscillations do indeed become smaller for
smaller $\Omega_\Lambda$, showing that they are a consequence
of the presence of the cosmological constant.

\section{Conclusion}

In this paper we have investigated the effects of the presence of a cosmological
constant in the contracting phase of a bouncing model. It turns out that the initial
vacuum state, usually determined in the
contracting phase
when the Universe was very large and rarefied, is affected by the
presence of the cosmological constant. In order to get an almost scale
invariant spectrum, one still must have some phase with dust-like contraction, but
now the spectral index gets a red-tilted running and oscillations directly caused
by the cosmological constant. It is interesting to realize that bouncing models
allow such an important role to the cosmological constant in the physics of primordial
cosmological perturbations, which is not at all the case for always expanding models.
This opens a new area of research, which is to investigate the influence
of other models of dark energy on the primordial spectrum of bouncing models..
In other words, if the Universe had really bounced in the past, investigating
its primordial spectrum can yield information about dark energy.

There is also the question about the possibility of an enormous growth of
perturbation amplitudes in the deflationary contraction
in the far past of the model, as discussed in other contexts \cite{goldwirth}. 
Note, however, that the cosmological constant in our model is small and this almost 
de Sitter deflation will take place when the Universe was very large compared to the 
present Hubble radius, and for a fixed time interval. In conformal time, this
time interval can be estimated using Eq.~(\ref{conformal}) by saying that it should be smaller 
than $\eta_* + \eta_{\infty}$ given by this equation when $t\approx -1/\sqrt{\Lambda}$.
Taking the usual values $\Omega_{\Lambda}=0.7$ and $\Omega_{0w}=0.3$, one can see that
$\eta_* + \eta_{\infty}<1$. Note that for a cosmological constant dominated model ($\Omega_{0w}\approx 0$),
one would obtain $\eta_* + \eta_{\infty}>>1$. Let us examine the behaviour of the Bardeen potential in 
this phase. From Eq.~(\ref{bds}) we obtain that 

\begin{equation}
\label{bardeen}
\Phi \propto \frac{(\epsilon + p)a^2}{k^2 {\cal{H}}}\left(\frac{v}{z}\right)' \approx c_1 (\eta + \eta_{_\infty})^{3} + 
\frac{c_2}{k^2} (\eta+ \eta_{\infty})
\end{equation}
for $w=0$. Hence, as this almost de Sitter deflationary phase will not take long enough in conformal time, 
$\eta_* + \eta_{\infty}<1$, because of the smallness of the cosmological constant, perturbations will not grow 
alarmingly in this epoch. Now, once the Universe leaves this deflationary contraction to a non-deflationary contraction when
it is still very large, then it can be subjected to dissipation effects, as the ones discussed in Ref.~\cite{ppn4}, which could dissipate the existing inhomogeneities. Another approach to this problem should be to think in terms of the Anthropic Principle
and state that the Universe is composed fundamentally 
by a small cosmological constant (dark energy) and some matter content as the one used in our model (dark matter?). 
In many regions it will expand to de Sitter and it will freeze, in some it will contract inhomogeneously, and in a few homogeneous 
regions within one particle horizon size it may contract to make a bounce, where new particles (photons and baryons) will be created, and expand again to a Universe with some galaxies and stars where intelligent life can exist. The results of our paper can then be 
applied to this last possibility, the only one which can interest us. 
Of course these tentative answers to this basic question must be worked out more precisely, but we think that 
a final and complete answer to the issue on why the primordial Universe was so homogeneous, in any cosmological 
scenario, demands a theory of initial conditions, perhaps quantum cosmology. One interesting investigation 
should be a quantum cosmological analysis of eternal asymptotically (in time) de Sitter models.

Our approach here was concentrated on general features of the spectrum
of cosmological perturbations in the presence of a cosmological constant in general
bouncing models, and because of that we were not able to fix the amplitude of the perturbations.
Our next step will be to take a specific model in which
the physics at the bounce fixes the bounce scale, and hence the amplitude of the
perturbations, in order to determine the influence of
the new fetures of the spectrum of primordial perturbations we obtained in
this paper in the anisotropies
of the microwave background radiation, and to compare the results with
observations. 
Another interesting problem should be to investigate the
situation where the required dust-like contraction was not caused by a scalar field
but by a hidrodynamical fluid with $c_s^2 = w$. In this case, the prescription
of an adiabatic vacuum can be much more involved because of the smallness
of the sound horizon.

\section*{Acknowledgements} RM, NPN and SP would like to thank CNPq of Brazil for
financial support. BBS would like to thank a PCI fellowship for financial support and Patrick Peter for his help with the numerical calculations.
NPN would also like to thank Slava Mukhanov and group of "Pequeno Semin\'ario" for
many useful comments and suggestions.

\end{document}